\documentclass[5p,twocolumn,number]{elsarticle}

\usepackage{graphics}
\usepackage{epsfig}
\usepackage{amssymb}

\begin{document}

\begin{frontmatter}



\title{QCALT: a tile calorimeter for KLOE-2 experiment}


\author{M.Cordelli$^a$, G.Corradi$^a$,F.Happacher$^a$,
        M.Martini$^{a,b,}$\footnote{Corresponding author: matteo.martini@lnf.infn.it},
        S.Miscetti$^{a}$,C.Paglia$^a$,A.Saputi$^a$,
        I.Sarra$^{a}$,D.Tagnani$^a$}

\address{$^a$ Laboratori Nazionali di Frascati dell'INFN, Frascati (Rm), Italy\\
         $^b$ Dipartimento di Energetica Univ. Roma La Sapienza, Roma, Italy.}

\begin{abstract}
The upgrade of the DA$\Phi$NE machine layout requires a modification
of the size and position of the inner focusing quadrupoles of 
KLOE-2 thus asking for  the realization of two new calorimeters 
covering the quadrupoles area.
To improve the reconstruction of $K_L\to 2\pi^0$ events 
with photons hitting the quadrupoles a calorimeter 
with high efficiency to low energy photons (20-300 MeV),
time  resolution of less than 1 ns  and space resolution of few cm,
is needed. 
To match these requirements, we are designing a tile
calorimeter, QCALT, where each single tile is readout
by mean of SiPM for a total granularity of 2400 channels.
We show first tests of the different calorimeter components.

\end{abstract}

\begin{keyword}
Kloe-2 \sep Calorimeters \sep SiPM \sep tiles \sep WLS fibers.



\end{keyword}

\end{frontmatter}


\section{The KLOE-2 proposal}

In the last decade a wide experimental program has
been carried out at DA$\Phi$NE\cite{dafne},
the $e^+e^-$
collider of the Frascati National Laboratories,
running at a center of mass energy of the 
$\phi$ resonance. During KLOE\cite{kloe_all} run, DA$\Phi$NE
delivered a peak luminosity of 1.5$\times$10$^{32}$
cm$^{-2}$s$^{-1}$ (corresponding to 1 fb$^{-1}$ per year). 

A new machine scheme\cite{crab} has been recently proposed 
to increase
the luminosity of the machine up to a factor 5.
This scheme has been succesfully tested at DA$\Phi$NE,
and the encouraging results pushed for 
a new data taking campaign.
This new phase will start a new experiment,
named KLOE-2, aiming to complete KLOE physics program.

To 
improve the performances
of the detector we expect to add
new subdetector systems such as: an inner tracker, a tagger
system to study $\gamma\gamma$ physics, a new small angle calorimeter
and a new quadrupole
calorimeter.
In this paper we explain 
the project and R\&D for this last detector.

\section{The quadrupole tile calorimeter, QCALT}

In the old IP scheme of DA$\Phi$NE 
the inner focalizing quadrupoles have two 
surrounding
calorimeters QCAL \cite{oldqcal}
covering a polar angle 
down to 21 degrees.
Each calorimeter consists of  16 azimuthal sectors composed
by alternating layers of 2 mm lead and 1 mm BC408 
scintillator tiles, for a total thickness of $\sim$5 X$_0$.
The fiber arrangement (back bending) allows
the measurement of the longitudinal coordinate by time differences
with a resolution of 13 cm.
These calorimeters are characterized by a low light response 
(1-3 pe/mip/tile)
due 
to the  coupling in air, to the fiber lenght 
($\sim$2 m for each tile) and 
to the quantum efficiency of the used photomultipliers (standard bialkali 
with $\sim$20\% QE).

\begin{figure}[htb]
\psfig{file=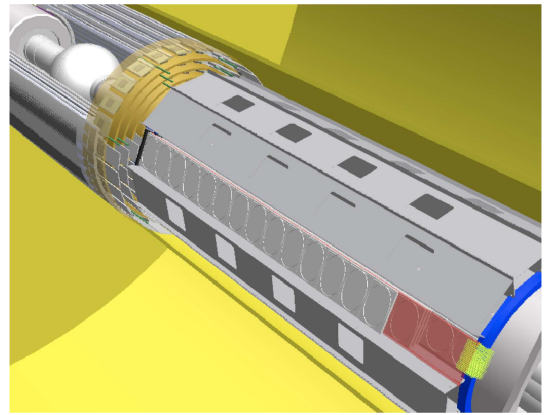,width=7cm}
\caption{IP scheme of KLOE-2. In this 
project are visible the inner tracker around IP and QCAL
sourronding quadrupoles.}
\label{figqcal}
\end{figure}

The project of the new QCAL 
(See Fig.\ref{figqcal})
consists in a dodecagonal structure,
1 m long, covering the region of
the new quadrupoles
composed by 
a sampling of 5 layers of 
5 mm thick scintillator plates alternated
with 3.5 mm thick tungsten plates,
for a total depth of 4.75 cm (5.5 X$_0$). 
The active part of each plane is divided into 
twenty tiles of $\sim$5$\times$5 cm$^2$ area 
with 1 mm diameter WLS fibers embedded 
in circular grooves. Each fiber is then
optically connected to a silicon 
photomultiplier of 1 mm$^2$ area, SiPM, 
for a total of 2400 channels.

We report the R\&D studies done on SiPM, fibers and tiles 
we have carried out to select the components
which optimize
the performance of our system.

\subsection{Tests performed on single components}

We have compared the characteristics of two different SiPM 
produced by Hamamatsu (multi pixel photon counter, MPPC):
100 (S10362-11-100U) and 400 pixels 
(S10362-11-050U), both with 1$\times 1$ mm$^2$ active area.

We have prepared a setup 
based on a blue light pulsed 
LED, a polaroid filter to modify the 
light intensity and a SiPM 
polarization/amplification circuit
based on Minicircuits MAR8-A+ amplifier.
We have measured the gain and the dark rate
variation as a function both of the applied V$_{bias}$
and the temperature of the photodetector.
The readout electronics was based on CAMAC, 
with a charge sensitivity of 0.25 pC/count
and a time of 125 ps/count. 

Our tests confirm the performances declared by Hamamatsu and
show a significative variation of the detector gain 
as a function of the temperature 
(3\% for 400 pixels versus 6\% for 100 pixels).

To decide the best fiber solution, 
we have studied the light response of two different, 
1 mm$^2$, WLS from blue to green, 
fibers optically connected to MPPC 
when hit by electrons produced by
a Sr$^{90}$ source: 
Saint Gobain BCF92 single cladding 
and
Saint Gobain BCF92 multi cladding.
The adopted solution is 
Saint Gobain BCF92 multi cladding.
For this fiber we 
find, as expected, a large light yield
than the one with single cladding
($\times 1.5$), a fast emission time 
(5 ns/pe) and long 
attenuation length.


Light response and time resolution of 
a complete tile have been 
measured using cosmic rays. 
The system was prepared connecting
fiber to MPPC and using two 
external NE110 scintillators fingers to trigger 
the signal. 
We have prepared different tiles (3 and 5 mm thick)
readout with 100 or 400 pixels MPPC. The adopted solution is 5 mm thick
BC408 tile readout by 400 pixels MPPC which balance 
the light yield optimization versus the dark rate. 
For this system we obtain 32 pe/mip with a time
resolution of 750 ps after correcting for the time dependence 
on pulse height. 

\begin{figure}[htb]
\vspace{9pt}
\psfig{file=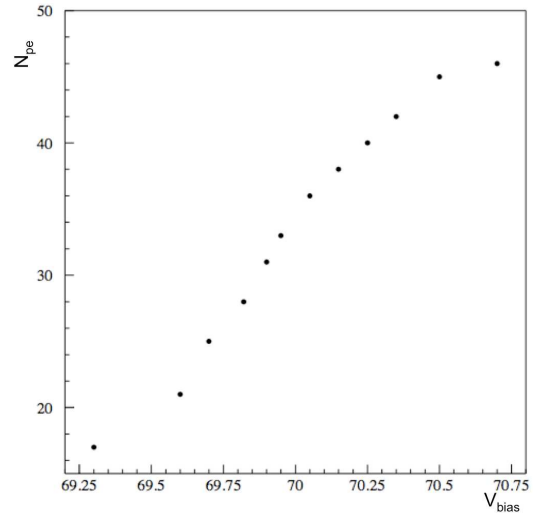,width=7.cm}
\caption{Signal detected by SiPM using a fixed LED
light and varying V$_{bias}$.}
\label{npevaria}
\end{figure}

Controlling enviromental conditions and using LED light, we 
have also studied SiPM response when varying 
V$_{bias}$.
By using the photon counting properties
of the SiPM we observe an increase of the light yield
when increasing V$_{bias}$ as shown in 
Fig.\ref{npevaria}.
The device reach 
a plateau 600 mV above operatin voltage, which is
consistent with a variation of the 
photon detection efficiency of the SiPM for the avalanche 
probability.

\subsection{FEE electronic}

To manage the signals for many channels,
the electronic 
service of the Frascati Laboratory has 
developed some custom electronics composed by
a 1$\times$2 cm$^2$ chip, containing the pre-amplifier
and the voltage regulator, and 
a multifunction NIM board. 
The NIM board 
supplies the V$_{bias}$ to the photodetector with 
a precision of 2 mV and a stability at the level of 
0.03 permill. A low threshold 
discriminator and a fanout are also present. 

\subsection{Next plans}

We are now assembling two small 
dimension 
prototypes of the QCAL (two full planes with 
20 tiles/each and one 
full column with 5 planes) , to study
both the signal transportation 
to the end and to 
measure the effective radiation length
at electron beam. 
For the end of 2009,
we plan also to construct a ``module 0''
consisting of a complete slice of the dodecagon 
(1/12$^{th}$ of one calorimeter) 
with final material and electronics. 

\section{Conclusions}
The new scheme proposed for DA$\Phi$NE machine allows
a factor 5 increase in the delivered luminosity. 
Some R\&D are in progress to add new 
components to the KLOE apparatus. 
We have presented the project and the 
R\&D in progress for
a tile 
calorimeter surronding the focalizing
quadrupoles.
Using 5$\times$5$\times$0.5 cm$^3$ BC408 tile,
readout by 400 pixels 
MPPC, we obtain 32 pe/mip with a time resolution
of 750 ps. 
\section{Acknowledgement}
We really thank 
M.Arpaia, G.Bisogni, A.Cassar\`a, A.Di Virgilio, U.Martini
and A.Olivieri for their help in the mechanical preparation
of the setup and in the 
preparation of the tiles.



\end{document}